\documentclass[
aps,%
12pt,%
final,%
notitlepage,%
oneside,%
onecolumn,%
nobibnotes,%
nofootinbib,%
superscriptaddress,%
noshowpacs,%
centertags]%
{revtex4}

\usepackage[english]{babel}
\usepackage[utf8]{inputenc}
\usepackage[T1]{fontenc}
\usepackage{amsmath}
\usepackage{amssymb}
\usepackage{graphicx}
\usepackage{hyperref}
\usepackage{color}
\usepackage{bm}
\usepackage{braket}

\begin{document}

\title{Weak Interaction Contribution to the Muonium Hyperfine Structure in the Standard Model}
\author{\firstname{F.~A.} \surname{Martynenko}}
\email{f.a.martynenko@gmail.com}
\affiliation{Samara University, Samara, Russia}
\author{\firstname{A.~P.} \surname{Martynenko}}
\affiliation{Samara University, Samara, Russia}
\author{\firstname{K.~A.} \surname{Seredina}}
\affiliation{Samara University, Samara, Russia}

\begin{abstract}
The contribution of the weak interaction to the hyperfine splitting of the ground state of muonium is investigated. 
The amplitudes of the one- and two-quantum exchange determined by the Z and W bosons are calculated. 
One-loop corrections in the photon and Z boson propagators and their contribution to the hyperfine structure 
of the spectrum are obtained.
\end{abstract}

\maketitle

\section{Introduction}

Muonium is a purely leptonic atom, devoid of internal structure. For many decades, this system has been and remains 
one of the most suitable for testing quantum electrodynamics, the Standard Model (SM), and searching for physics 
beyond the Standard Model. Precision studies of low-lying muonium energy levels have focused on a number 
of energy intervals \cite{paper11}.
The study of the hyperfine structure (HFS) of the ground state of muonium is one of the central tasks in the physics 
of the simplest bound states of particles. Spin-dependent interactions between the electron and muon arising 
from boson exchange have been investigated in various experiments in the low-energy region.
The current theoretical prediction of hyperfine splitting of the ground state in muonium is \cite{paper1}:
\begin{equation}
\label{d1}
\Delta E^{hfs}_{th}=4\,463\,302\,872(515)~\text{Hz}, \quad \delta=1.3\cdot 10^{-7},
\end{equation}
where the largest part of the uncertainty (511 Hz) is due to the measurement of the lepton mass ratio $m_\mu/m_e$ (120 ppb). 
In \eqref{d1} and below, the energy interval is expressed in Hz according to the formula $\Delta E=h\Delta\nu$. 
A more precise value of the mass ratio $m_\mu/m_e$ can be obtained from a comparison of theoretical and new 
experimental results for the hyperfine splitting of muonium atoms. The goal of the MuSEUM collaboration 
is to precisely measure the ground-state hyperfine splitting of muonium atoms with an accuracy of 1 ppb \cite{paper2}.
At the J-PARC muon science facility (MUSE), the MuSEUM collaboration is conducting new precision measurements 
of the hyperfine structure (HFS) of the ground state of muonium and muonic helium. 
It is precisely high-precision measurements of the hyperfine structure of the ground state of muonium, along with the anomalous magnetic moment of the muon, that are the most sensitive tool for testing the quantum electrodynamics 
of bound states and the Standard Model as a whole.


It should also be noted that there are several other energy intervals for muonium being investigated experimentally 
with high precision. The latest experimental result of measurement of the transition frequency $(1S\div 2S)$ 
in muonium was obtained in \cite{paper3}:
\begin{equation}
\label{d2}
\Delta E^{exp}(1S-2S)= 2\,455\,528\,941.0(9.8)~\text{MHz}.
\end{equation}
The muonium laser spectroscopy collaboration (the Mu-MASS experiment) plans to measure the muonium $(1S-2S)$ 
transition frequency with a final uncertainty of 10 kHz, providing a thousandfold improvement in accuracy \cite{paper4,paper5}.

A new measurement result for the Lamb shift for $n=2$ in muonium represents an order of magnitude improvement 
in accuracy compared to the previous best measurement \cite{paper6}:
\begin{equation}
\label{d3}
\Delta E^{exp}(2S_{1/2}-2P_{1/2})=1047.2 ~(2.3)_{\text{stat}} ~(1.1)_{\text{syst}} ~\text{MHz}.
\end{equation}
The theoretical prediction for the Lamb shift in a muonium is the following \cite{paper7}
\begin{equation}
\label{d4}
\Delta E^{th}(2S_{1/2}-2P_{1/2})=1047.284(2)~\text{MHz},
\end{equation}
hence, calculating various contributions on the order of 1 kHz is currently a very relevant task.

A measurement of the $2S_{1/2}-2P_{3/2}$ fine-structure transition in muonium using microwave spectroscopy 
gave the transition frequency \cite{paper8}
\begin{equation}
\label{d5}
\Delta E^{exp}(2S_{1/2}-2P_{3/2})=9871.0\pm 7.0~\text{MHz},
\end{equation}
which agrees with modern quantum electrodynamics predictions of $9874.367 \pm 0.001$ MHz \cite{paper9} 
within one standard deviation.

The last experimental measurement of hyperfine structure (HFS) in muonium was performed more than 
twenty-five years ago with an accuracy of up to the hundredths of a kHz \cite{paper10,paper10a,paper11}:
\begin{equation}
\label{d6}
\Delta E^{hfs}_{exp}(1S)=4\,463\,302\,765 (53)~\text{Hz}.
\end{equation}

Theoretical studies of the hyperfine muonium structure have been conducted for a long time and have reached 
very high accuracy \cite{paper1}.
The accuracy of calculating purely quantum-electrodynamic contributions is already determined by corrections of the 7th-8th order in the fine-structure constant. However, in conjunction with such contributions, contributions 
determined by other interactions, strong and weak, may become important for muonium HFS \cite{ibk,paper12,paper12a,feinberg,fs,eides,miv,weak1,weak2,alcorta,weak3,weak4,av}.
Electroweak corrections are suppressed for fine and hyperfine structure intervals due to the appearance of the 
$W^\pm$ and $Z$ boson masses in the denominators of the corresponding energy intervals. Consequently, they have not received much attention, for example, in studies of the hyperfine structure of muonium. Nevertheless, the situation 
with respect to the calculation of electroweak contributions may currently change qualitatively, driven 
by a significant increase in experimental accuracy for energy intervals in muonium and more intensive investigations in the Higgs sector after the discovery of the Higgs boson \cite{weak1,weak2,alcorta,weak3,weak4,weak5a,weak5,weak6,weak7}.
Thus, current theoretical calculations involve investigating new previously unaccounted interactions and calculating 
new contributions to the HFS considering the planned experimental accuracy.

There are different types of electroweak corrections to interaction amplitudes. These primarily include one-loop corrections to the photon and Z-boson propagation functions, as well as one-loop box-type amplitudes with $Z$ and $W$ bosons.  We consider precisely such contributions to obtain numerical estimates of the electroweak interaction in the muonium hyperfine structure.

\section{Contribution of One Z Boson Exchange}

\begin{figure}[htbp]
\centering
\includegraphics[scale=1.2]{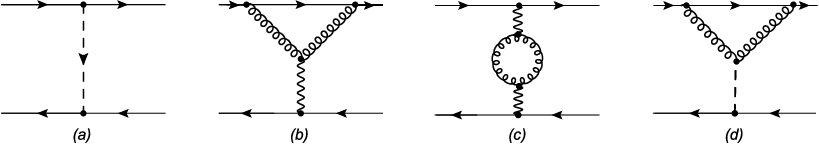}
\caption{One-boson interaction amplitude and one-photon interaction with vertex and self-energy corrections. 
The dashed line denotes the $Z$ boson, the wavy line denotes the photon and the curlicue line denotes the W-boson.}
\label{fig1}
\end{figure}

The effects of a weak interaction are determined by the exchange of heavy charged gauge bosons $W^\pm$ and neutral 
bosons $Z$.
The main contribution of the weak interaction to the muonium HFS is determined by the amplitude of the one Z boson interaction shown in Fig.~\ref{fig1}(a) \cite{feinberg,fs,eides}.

To construct this amplitude, we take into account that the vertex function for the interaction of a $Z$ boson 
with a lepton has the following form \cite{CORE}:
\begin{equation}
\label{f1}
\Gamma^\alpha=G_V^f\gamma^\alpha+G_A^f\gamma^\alpha\gamma_5=
\frac{ie}{\sin 2\theta_W}\gamma^\alpha\left[2T_f^3\frac{(1-\gamma_5)}{2}-2Q_f\sin^2 \theta_W\right],
\end{equation}
where $T^3_f$ and $Q_f$ denote the third component of the weak isospin and the charge of the fermion, 
respectively.
For example, for electron $T_e^3=-\frac{1}{2}$, $Q_e=-1$. For heavy quarks, their contributions with fermion 
loops are considered in the following $T_{c,t}^3=\frac{1}{2}$, $Q_c=\frac{2}{3}$, $T_{b}^3=-\frac{1}{2}$, $Q_b=-\frac{1}{3}$.
Note that the interaction of the weak neutral current \eqref{f1} remains unchanged when going from particle 
to antiparticle \cite{eides}. For weak mixing angle $\theta_W$ we take the value from \cite{CODATA}:
$\sin^2\theta_W(M_Z^2)=1-\frac{M_W^2}{M_Z^2}=0.22305(23)$.

We choose the $Z$-boson propagator in the Coulomb gauge, which is best suited for bound-state calculations 
and is described in detail in \cite{p1,p2,p2a,p3}:
\begin{equation}
\label{f2}
D^{\mu\nu}(k)=\frac{1}{(k^2-M_Z^2)}\left[
g^{\mu\nu}+\frac{1}{2}\frac{1}{({\bf k}^2+M_Z^2)}(f^\mu k^\nu+f^\nu k^\mu)\right],\quad f^0=-k^0,\quad f^i=k^i,
\end{equation}
where k is the 4-momentum of the $Z$ boson.

The hyperfine part of the potential is determined by bispinor contractions, which have the following form 
for the first and second particles \cite{blp}:
\begin{eqnarray}
\label{f3}
\bar u'_1\boldsymbol\gamma_1 u_1&=&\mathbf{w}'^\ast_1\left\{
i[\boldsymbol\sigma_1\times{\bf k}]+2{\bf p}_1+{\bf k}\right\}\mathbf{w}_1,\\
\label{f3a}
\bar u'_2\boldsymbol\gamma_2 u_2&=&\mathbf{w}'^\ast_2\left\{
-i[\boldsymbol\sigma_2\times{\bf k}]+2{\bf p}_2-{\bf k}\right\}\mathbf{w}_2,\\
\label{f3b}
\bar u'_1\gamma_0\gamma_5 u_1&=&\mathbf{w}'^\ast_1\left\{(\boldsymbol\sigma_1{\bf p}_1)+
(\boldsymbol\sigma_1{\bf p}'_1)\right\}\mathbf{w}_1,\\
\label{f3c}
\bar u'_2\gamma_0\gamma_5 u_2&=&\mathbf{w}'^\ast_2\left\{(\boldsymbol\sigma_2{\bf p}_2)+
(\boldsymbol\sigma_2{\bf p}'_2)\right\}\mathbf{w}_2,
\end{eqnarray}
where $\mathbf{w}_i$, $\mathbf{w}'_i$ ($i=1,2$) are the spinor wave functions of the first and second particles 
in the initial and final states. In the initial state, the particle momenta are ${\bf p}_1$, ${\bf p}_2$, 
and in the final state they are ${\bf p}'_1$, ${\bf p}'_2$.

Then the vector part of the neutral weak current gives the following contribution to the interaction potential in momentum representation:
\begin{equation}
\label{f4}
\Delta U_V^{hfs}({\bf k})=-\frac{G^2_V}{6m_1m_2}\left[1-\frac{M_Z^2}{{\bf k}^2+M_Z^2}
\right](\boldsymbol\sigma_1\boldsymbol\sigma_2),
\end{equation}
where $m_1$, $m_2$ are the electron and muon masses.

The pseudovector part of the neutral weak current also contributes to the interaction potential in momentum representation as follows \cite{feinberg,fs,eides}:
\begin{equation}
\label{f4a}
\Delta U_{AV}^{hfs}({\bf k})=-\frac{G_A^2}{({\bf k}^2+M_Z^2)}\left[1-\frac{1}{3}\frac{{\bf k}^2}{{\bf k}^2+M_Z^2}
\right](\boldsymbol\sigma_1\boldsymbol\sigma_2).
\end{equation}

Using the Fourier transform, we obtain the corresponding hyperfine interaction potentials in coordinate 
representation:
\begin{equation}
\label{f5}
\Delta U_{V}^{hfs}(r)=-\frac{2G_V^2}{3m_1m_2}\left[\delta({\bf r})-
\frac{M_Z^2}{4\pi r}e^{-M_Zr}\right]
\frac{(\boldsymbol\sigma_1\boldsymbol\sigma_2)}{4}, 
\Delta U_{AV}^{hfs}(r)=-\frac{2G_A^2}{3\pi r}(1+\frac{M_Z r}{4})e^{-M_Zr}
\frac{(\boldsymbol\sigma_1\boldsymbol\sigma_2)}{4}.
\end{equation}

Their contributions to the hyperfine splitting of the muonium ground state are determined after averaging the wave function by the following expressions:
\begin{eqnarray}
\label{f5a}
\Delta E^{hfs}_{AV}&=&-\frac{G_F\sqrt{2}\mu^3\alpha^3}{\pi}(1-\frac{7\mu\alpha}{3M_Z})=-64.8827~\text{Hz},\quad
E_F=\frac{8\mu^3\alpha^4}{3m_1m_2},\\
\label{f5b}
\Delta E^{hfs}_{V}&=&-\frac{\mu M_Z G_F}{2\sqrt{2}\pi}(1-4\sin^2\theta_W)^2 E_F=-3.1505~\text{Hz},
\end{eqnarray}
where $\mu=m_1m_2/(m_1+m_2)$.
Numerical values are presented in Hz with up to four decimal places. Despite contributions \eqref{f5a}, \eqref{f5b} 
being of the order $\alpha^4$, $\alpha^5$, they contain additional mass factors that lead to a significant reduction 
in contribution.
In the limit $M_Z\to\infty$, the potential 
$\Delta U^{hfs}_{AV}({\bf k})$ reduces to a constant 
and the potential $\Delta U^{hfs}_{AV}(r)$ to the Dirac
$\delta$-function: $\Delta U^{hfs}_{AV}(r)=-\frac{4G_A^2}{M_Z^2}\frac{(\boldsymbol\sigma_1\boldsymbol\sigma_2)}{4}\delta({\bf r})$.
The correction $\mu\alpha/M_z$ in \eqref{f5a} is negligible.
It should be noted that, following work \cite{eides}, we expressed the results \eqref{f5a}-\eqref{f5b} through 
the Fermi constant
$G_F=\frac{\pi\alpha(M_Z^2)}{\sqrt{2}\sin^2\theta_W(M_Z^2) M_W^2}$
\cite{miv} for which the following value was used \cite{CODATA}: $G_F=1.166378\times 10^{-5}~\text{GeV}^{-2}$,
$\alpha^{-1}(M_Z^2)= 128.947\pm 0.013$ \cite{davier}.
The numerical values \eqref{f5a}-\eqref{f5b} show that one-loop corrections to the photon and $Z$-boson propagator functions may be significant considering the growth in experimental accuracy of HFS measurements.

Calculations show that in the second order of perturbation theory (sopt), contributions to the hyperfine structure are suppressed by the appearance of an additional power of the parameter $\mu\alpha/M_Z$. For example, in the case 
of two hyperfine perturbation potentials, the contribution to the hyperfine structure is:
\begin{equation}
\label{f5c}
\Delta E_{sopt}^{hfs}=2\langle\psi|\Delta U^{hfs}_{AV}(\tilde G-G^f)\Delta V^{hfs}_\gamma|\psi\rangle=
E_F\frac{\mu^2\alpha^2}{3\pi\sin^2 2\theta_W M_Z^2}\left(\frac{5}{2}\ln\frac{\mu\alpha}{M_Z}-\frac{7}{2}\right)=-0.0001~\text{Hz},
\end{equation}
where $\tilde G$ is the reduced Coulomb Green's function,
$G^f$ is the free Coulomb Green's function, $\Delta U^{hfs}_\gamma$ is the hyperfine part of the $1\gamma$-potential \cite{blp}.

\begin{figure}[htbp]
\centering
\includegraphics[scale=1.]{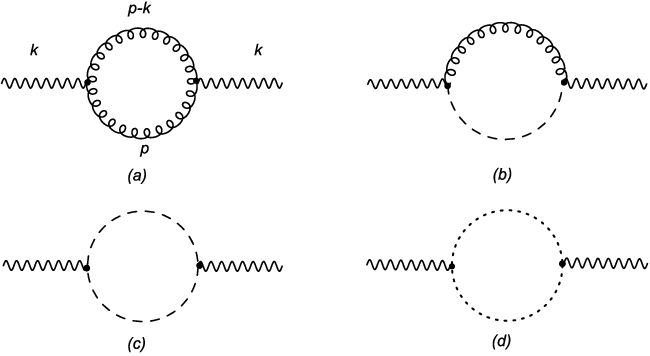}
\caption{Self-energy correction to the photon propagator function with loops of $W$ bosons -- (a,b), gauge bosons -- (b,c), and ghosts -- (d). The wavy line denotes the photon. The dashed line denotes the pseudo-Goldstone boson 
$\omega$, and the dotted line denotes the ghost field.}
\label{fig1a}
\end{figure}

In the one-photon interaction, there is a self-energy correction to the HFS potential determined by a loop of $W$ 
bosons (Fig.~\ref{fig1a}(a)).
Alongside this loop, in an arbitrary $R_{\xi}$ gauge, it is necessary to consider loop contributions from pseudo-Goldstone bosons and ghosts. The contributions of the corresponding loops are shown in Fig.~\ref{fig1a}(b,c,d). 
Usually, to avoid accounting for contributions from unphysical gauge bosons and ghosts, one chooses the unitary 
gauge, where the gauge parameter $\xi=\infty$.
The $\xi$-independence of physical process amplitudes to all orders of perturbation theory in non-Abelian gauge 
theories with the Higgs mechanism was proven by Veltman, 't Hooft, and others \cite{miv}.
For any finite $\xi$, the particle propagators fall off as $\frac{1}{k^2}$, i.e., give a renormalizable theory. 

Subsequently, when one-loop amplitudes are calculated by the dispersion method, it is convenient to use the 't Hooft-Feynman gauge ($\xi=1$). In this gauge, the form of the particle propagators in the loops simplifies significantly.
Dispersion relations are a reliable tool for calculating radiative corrections in quantum electrodynamics and quantum field theory, in general. The general idea of this method is that the structure functions of the particle interaction amplitude can be calculated using their imaginary parts, which, in turn, can be obtained in final form by directly calculating the Feynman integrals \cite{nussenzveig,nishijima}.

The vertex for the interaction of a photon with a pair of $W$ bosons in Fig.~\ref{fig1a}(a) has the form:
\begin{equation}
\label{f6}
\Gamma^{\mu\beta\gamma}=(-ie)\left[g^{\gamma\beta}(k-2p)^\mu+g^{\beta\mu}(p-2k)^\gamma+
g^{\gamma\mu}(k+p)^\beta\right].
\end{equation}

In the 't Hooft-Feynman gauge, the contributions to the polarization operator from various particles (bosons $W^\pm$, pseudo-Goldstone bosons $\omega^\pm$, ghosts $c^\pm$) in the loops have the following structure:
\begin{equation}
\label{f6a}
P^{\mu\nu}(k)=\left(g^{\mu\nu}-\frac{k^\mu k^\nu}{k^2}\right)A(k^2)+
\frac{k^\mu k^\nu}{k^2}B(k^2),
\end{equation}
i.e., contains transverse and longitudinal parts. Considering first the contribution of the $W$-boson loop, denoted 
below by the index $WW$ for various functions, we note that after contraction over Lorentz indices in the numerator 
of the polarization tensor $P^{\mu\nu}_{WW}(t)$, one can isolate, using the Form package \cite{Form}, two parts corresponding to $A_{WW}$, $B_{WW}$, in the form (in the frame where the 4-momentum $k=(k_0,0)$):
\begin{equation}
\label{f7}
{\cal N_{A~WW}}=2k_0p_0-5k_0^2+\frac{10}{3}p_0^2-\frac{16}{3}p^2,
\end{equation}
\begin{equation}
\label{f7a}
{\cal N_{B~WW}}=12k_0p_0-3k_0^2-10p_0^2-2p^2.
\end{equation}

To calculate the functions $A_{WW}(t)$ and $B_{WW}(t)$ themselves, we use the dispersion renormalization method \cite{blp}. When calculating the imaginary part of the amplitude in accordance with the Mandelstam-Cutkosky rule, 
we make the following substitutions in the W-boson propagators:
\begin{equation}
\label{f9}
\frac{1}{(p^2-M_W^2)}\to (-2\pi i)\delta_+(p^2-M_W^2),\quad \frac{1}{(p-k)^2-M_W^2}\to (-2\pi i)\delta_+((p-k)^2-M_W^2).
\end{equation}

Then the product of two $\delta$-functions can be transformed as follows:
\begin{equation}
\label{f10}
(-2\pi i)^2\delta_+(p^2-M_W^2)\delta_+((p-k)^2-M_W^2)=-4\pi^2\frac{1}{2k_0}\delta\left(p_0-\frac{k_0}{2}\right)
\frac{1}{2\varepsilon(p)}\delta(\varepsilon-p_0),
\end{equation}
and the imaginary part of the functions $A_{WW}(t)$, $B_{WW}(t)$ takes the form:
\begin{equation}
\label{f11}
\text{Im} A_{WW}(t)=\frac{\alpha}{12}\sqrt{\frac{t-4M_W^2}{t}}\left(\frac{19}{2}t+16M_W^2\right), 
\text{Im} B_{WW}(t)=-\frac{\alpha}{8}\sqrt{\frac{t-4M_W^2}{t}}\left(t-4M_W^2\right).
\end{equation}

The imaginary parts of the structure functions for loops with gauge bosons $\omega^\pm$ and $W^\pm$ are calculated similarly. The vertices of the particle interactions in the amplitudes of Fig.~\ref{fig1a}(b)-\ref{fig1a}(c) are determined by the factors \cite{CORE}:
\begin{equation}
\label{f11a}
V^{\mu\beta}_{\gamma W\omega}=(-ie)M_W g^{\mu\beta},\quad V^\mu_{\gamma\omega\omega}=(-ie)(2p-k)^\mu.
\end{equation}

As a result of calculating the imaginary parts, the following expressions are obtained:
\begin{equation}
\label{f11b}
\text{Im} A_{W\omega}(t)=-\frac{\alpha}{2}\sqrt{\frac{t-4M_W^2}{t}}M_W^2,\quad
\text{Im} B_{W\omega}(t)=-\frac{\alpha}{2}\sqrt{\frac{t-4M_W^2}{t}}M_W^2,
\end{equation}
\begin{equation}
\label{f11c}
\text{Im} A_{\omega\omega}(t)=-\frac{\alpha}{12}\sqrt{\frac{t-4M_W^2}{t}}(t-4M_W^2),\quad
\text{Im} B_{\omega\omega}(t)=0.
\end{equation}

The last in this chain of amplitudes is the loop with ghost fields $(c)$, for which contributions to the structure functions $\text{Im} A(t)$, $\text{Im} B(t)$ are obtained as:
\begin{equation}
\label{f11d}
\text{Im} A_{cc}(t)=\frac{\alpha}{2}\sqrt{\frac{t-4M_W^2}{t}}\left(\frac{1}{12}t-\frac{1}{3}M_W^2\right),\quad
\text{Im} B_{cc}(t)=\frac{\alpha}{2}\sqrt{\frac{t-4M_W^2}{t}}\frac{1}{4}t.
\end{equation}

The total contribution of all four loops \eqref{f11}, \eqref{f11b}, \eqref{f11c}, \eqref{f11d} to the structure 
functions $\text{Im} A_{\gamma\gamma}$, $\text{Im} B_{\gamma\gamma}$, denoted by a single index $\gamma\gamma$, 
is the following:
\begin{equation}
\label{f11e}
\text{Im} A_{\gamma\gamma}(t)=\frac{\alpha}{2}\sqrt{\frac{t-4M_W^2}{t}}\left(\frac{3}{2}t+2M_W^2\right),\quad
\text{Im} B_{\gamma\gamma}(t)=0,
\end{equation}
which means that the polarization operator is purely transverse.

Using the standard dispersion relation for the polarization operator with two subtractions when constructing the renormalized expression, we can represent the structure function $A_{\gamma\gamma}(t)$ in integral form \cite{blp,nussenzveig}:
\begin{equation}
\label{f11f}
A_{\gamma\gamma}(t)=\frac{\alpha}{2\pi}t^2\int_1^\infty\frac{(3+\frac{1}{\xi^2})\sqrt{\xi^2-1}d\xi}{\xi^2({\bf k}^2+4M_W^2\xi^2)},
\end{equation}
where, in the case of bound states, the square of the transferred momentum is $t=k^2=-{\bf k}^2$ and is substituted explicitly into the denominator of the dispersion integral \eqref{f11f}. This integral relation is convenient for constructing the particle interaction operator in the coordinate representation and calculating corrections 
to the energy levels.

It follows from \eqref{f11f} that the contribution to the spin-independent part of the particle interaction operator 
in muonium from this vacuum polarization effect is given by:
\begin{equation}
\label{f12}
\Delta U_{\gamma\gamma}(r)=-\frac{\alpha}{2\pi}\int_1^\infty \rho_{\gamma\gamma}(\xi)\left(
-\frac{Z\alpha}{r}e^{-2M_W\xi r}\right)d\xi,\quad \rho_{\gamma\gamma}(\xi)=\frac{\sqrt{\xi^2-1}}{\xi^4}(3\xi^2+1).
\end{equation}

When constructing the potential in the momentum representation, one can set with necessary accuracy ${\bf k}=0$ ($t=0$) in \eqref{f11f} and calculate the integral over $\xi$:
\begin{equation}
\label{f12a}
\Delta U_{\gamma\gamma}(0)=-\frac{17\alpha(Z\alpha)}{30 M_W^2}.
\end{equation}

In coordinate representation, this potential is proportional to the Dirac $\delta$-function, and its contribution 
to the Lamb shift is:
\begin{equation}
\label{f13}
\Delta U_{\gamma\gamma}(r)=-\frac{17\alpha(Z\alpha)}{30 M_W^2}\delta({\bf r}),\quad
\Delta E_{\gamma\gamma}=-\frac{17\alpha(Z\alpha)^4\mu^3}{30\pi M_W^2}=-0.0184~\text{Hz},
\end{equation}
where the numerical value for the $n=1$ level shift is given.

We are interested in the contribution to the potential of the hyperfine structure of the muonium spectrum, which is constructed in this case in the same way as the contribution of the effect of electron vacuum polarization \cite{apm2005,apm2004}:
\begin{equation}
\label{f14}
\Delta U^{hfs}_{\gamma\gamma}(r)=-\frac{8\alpha}{3m_1m_2}\frac{\alpha}{2\pi}({\bf s}_1{\bf s}_2)
\int_1^\infty\rho_{\gamma\gamma}(\xi)\Bigl[\pi\delta({\bf r})-\frac{M_W^2\xi^2}{r}e^{-2M_W\xi r}
\Bigr]d\xi.
\end{equation}

Calculating the coordinate integral analytically with the $1S$-state wave function and using an expansion in the small parameter $\mu\alpha/M_W$, we represent the corresponding correction to the ground-state hyperfine splitting as:
\begin{equation}
\label{f15}
\Delta E^{hfs}_{\gamma\gamma}=-E_F\frac{\mu\alpha^2}{2\pi M_W}\left(\frac{13\pi}{8}-
\frac{17\mu\alpha}{5M_W}\right)=-1.2192~\text{Hz}.
\end{equation}

\begin{figure}[htbp]
\centering
\includegraphics[scale=1.3]{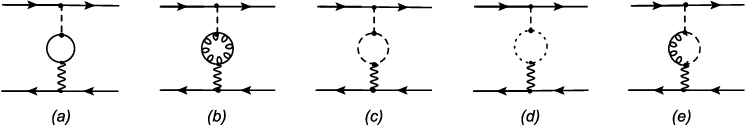}
\caption{One-loop corrections in the $Z\gamma$ interaction amplitude: fermion loop -- (a), $W^\pm$-boson loop -- (b), loops of pseudo-Goldstone bosons $\omega^\pm$ and ghosts -- (c), (d), loop of $W^\pm$ bosons and 
pseudo-Goldstone bosons $\omega^\pm$ -- (e). The wavy line denotes the photon, and the dashed line denotes 
the $Z$ boson.}
\label{fig1ae}
\end{figure}

The obtained numerical value of the contribution \eqref{f15} shows that it is necessary to consider the fermionic and bosonic self-energy corrections in the $\gamma Z$ propagator, presented in Fig.~\ref{fig1ae}. The peculiarity of these contributions is that one of the leptons or quarks interacts with the photon, and the other with the $Z$-boson.

Let us first consider the contribution of $Z\gamma$ interaction with a loop of leptons or quarks 
in Fig.~\ref{fig1ae}(a). The polarization operator is defined by the following expression:
\begin{equation}
\label{f16}
P^{\mu\nu}_{Z\gamma}(k)=-(-i)^4\frac{4\pi\alpha Q_f}{\sin 2\theta_W}\int\frac{d^4 p}{(2\pi)^4}
\text{Tr}\left\{\gamma^\mu \frac{(\hat p-\hat k+m_f)}{((p-k)^2-m_f^2)}\gamma^\nu
\left[2T^3_f\frac{(1-\gamma_5)}{2}-a_z\right]
\frac{(\hat p+m_f)}{(p^2-m_f^2)}\right\},
\end{equation}
where $m_f$ denotes the mass of the fermion in the loop, $a_z=2Q_f\sin^2\theta_W$.
Omitting the part of the trace with the $\gamma_5$ Dirac matrix and calculating the imaginary part 
of the polarization operator as in \eqref{f11}, we obtain the contribution from the fermion loop (f-loop):
\begin{equation}
\label{f17}
\text{Im} A_{f-loop}(t)=\frac{\alpha|Q_f|(1-2|Q_f|\sin^2\theta_W)N_c}{3\sin 2\theta_W}\sqrt{\frac{t-4m_f^2}{t}}(t+2m_f^2),~
\text{Im} B_{f-loop}(t)=0, ~ t\geq 4m_f^2.
\end{equation}
Formula \eqref{f17} is valid for both leptons and quarks: $|Q_c|=\frac{2}{3}$, $|Q_b|=\frac{1}{3}$, 
$|Q_t|=\frac{2}{3}$. The color factor $N_c=3$ for heavy quarks $c$, $b$, $t$.

To obtain the polarization operator in the $Z\gamma$ interaction, we use the dispersion relation with two 
subtractions ($P_{Z\gamma}(0)=0$, $P_{Z\gamma}(M_Z^2)=0$) \cite{nussenzveig}:
\begin{equation} 
\label{f18}
P_{Z\gamma}(t)=\frac{t(t-M_Z^2)}{\pi}\int_{4m_f^2}^\infty\frac{\text{Im} P_{Z\gamma}(t')dt'}
{(t'-t-i0)t'(t'-M_Z^2)},
\end{equation}
where the integral is understood in the sense of the principal value if $M_Z^2/4m_f^2 > 1$.
If we denote the unregularized integral,
\begin{equation}
\label{f18a}
\bar P_{Z\gamma}(t)=\frac{1}{\pi}\int_{4m^2}^\infty\frac{\text{Im} P_{Z\gamma}(t')dt'}{(t'-t-i0)},
\end{equation}
then the regularized polarization operator \eqref{f18} is obtained after subtractions of the form:
\begin{equation}
\label{f18b}
P_{Z\gamma}(t)=\bar P_{Z\gamma}(t)-\bar P_{Z\gamma}(0)-\frac{t}{M_Z^2}\left(\bar P_{Z\gamma}(M_Z^2)-
\bar P_{Z\gamma}(0)\right).
\end{equation}

The contribution to the spin-independent interaction potential in coordinate representation in leading order 
with respect to the ratio $m_f/M_Z$ is equal to
\begin{equation}
\label{f19}
\Delta U_{Z\gamma, f-loop}(r)=-\frac{8|Q_f|\alpha^2N_c(1-2|Q_f|\sin^2\theta_W)}{3M_Z^2\sin 2\theta_W}\left(
\ln\frac{2m_f}{M_Z}+\frac{5}{6}-\ln 2\right)\delta({\bf r}).
\end{equation}

The hyperfine part of the interaction operator is determined by the expression:
\begin{equation}
\label{f20}
\Delta U^{hfs}_{Z\gamma, f-loop}(r)=\frac{8|Q_f|\alpha N_c(1-2|Q_f|\sin^2
\theta_W)}{3m_1m_2\sin 2\theta_W}({\bf s}_1{\bf s}_2)\times
\end{equation}
\begin{displaymath}
\frac{\alpha}{3\pi}\int_1^\infty\rho_{Z\gamma}(\xi)d\xi\left(\pi\delta({\bf r})-
\frac{m_f^2\xi^2}{r}e^{-2m_f\xi r}\right),\quad
\rho_{Z\gamma}(\xi)=\frac{\sqrt{\xi^2-1}}{\xi^2}\frac{(\xi^2+\frac{1}{2})}
{(\xi^2-\kappa_1^2)},\quad \kappa_1=\frac{M_Z}{2m_f}.
\end{displaymath}

Other contributions to the hyperfine structure come from loops of leptons (electron, muon, tau-lepton) and heavy 
quarks (c, b, t). For the first five loops (lepton loops and $c$, $b$ quark loops),  the parameter 
$\kappa=M_Z/2m_f > 1$ ($m_f$ is the fermion mass in the loop) and the integral over $\xi$ in \eqref{f18} 
is understood in the sense of the principal value. For the $t$ quark $\kappa=M_Z/2m_f < 1$, so the integrand 
in the dispersion integral \eqref{f18} does not contain a singularity.
Analytical formulas for a calculation of contributions of fermion loops to the hyperfine structure 
of the spectrum take the form:
\begin{equation}
\label{f22}
\Delta E^{hfs}(e,\mu,\tau;c,b-\text{quarks})=-E_F\frac{|Q_f|\alpha^2\mu 
(1-2|Q_f|\sin^2\theta_W)}
{2m_f\sin 2\theta_W}\frac{(3\kappa_1^2+2)}{\kappa_1^4},
\end{equation}
\begin{equation}
\label{f23}
\Delta E^{hfs}(t-\text{quark})=E_F\frac{|Q_t|\alpha^2\mu(1-2|Q_f|\sin^2\theta_W)}{2m_t\sin 2\theta_W\kappa_1^4}\Bigl[
\kappa_1^2(4\sqrt{1-\kappa_1^2}-3)+2(\sqrt{1-\kappa_1^2}-1)\Bigr].
\end{equation}

As a result, the numerical values of the contributions \eqref{f22}, \eqref{f23} are:
\begin{equation}
\label{f23a}
\Delta E^{hfs}=
\begin{cases}
-0.00001~\text{Hz},\quad e-\text{loop},\\
-0.0020~\text{Hz},\quad \mu-\text{loop},\\
-0.0343~\text{Hz},\quad \tau-\text{loop},\\
-0.1140~\text{Hz},\quad c-\text{quark loop},\\
-0.4377~\text{Hz},\quad b-\text{quark loop},\\
-0.6768~\text{Hz},\quad t-\text{quark loop}.
\end{cases}
\end{equation}

The contribution of the other four loop amplitudes with $Z\gamma$ interaction in Fig.~\ref{fig1ae} is calculated similarly to \eqref{f11}-\eqref{f15}. The sum of these amplitudes in the polarization operator contains transverse 
and longitudinal parts:
\begin{equation}
\label{f24}
P^{\mu\nu}_{Z\gamma}=\left(g^{\mu\nu}-\frac{k^\mu k^\nu}{k^2}\right)A_{Z\gamma}(k^2)+
\frac{k^\mu k^\nu}{k^2}B_{Z\gamma}(k^2).
\end{equation}

It is well known that only the transverse component of the polarization operator contributes to $S$-matrix elements, since the longitudinal part vanishes when contracted with the polarization vector: $\varepsilon_\mu k^\mu=0$. Consequently, at some stage, the longitudinal component must cancel; otherwise, this condition would 
not be satisfied.

The one-loop $W$-boson amplitude in Fig.~\ref{fig1ae} is defined by the following integral expression:
\begin{equation}
\label{f25}
P^{\mu\nu}_{Z\gamma,W}=(-i)^2(-ie)^2 \cot{\theta_W}\int\frac{d^4p}{(2\pi)^4}
\left[(k-2p)_\mu g_{\gamma\beta}+(p-2k)_\gamma g_{\beta\mu}+(k+p)_\beta g_{\gamma\mu}\right]\times
\end{equation}
\begin{displaymath}
\left[(2p-k)_\nu g_{\lambda\rho}+(2k-p)_\rho g_{\lambda\nu}-(k+p)_\lambda g_{\rho\nu}\right]
\frac{g_{\beta\lambda}}{((p-k)^2-M_W^2)}\frac{g_{\gamma\rho}}{(p^2-M_W^2)}.
\end{displaymath}

Calculating the imaginary part of the polarization operator \eqref{f25}, we obtain for the contribution 
to its transverse and longitudinal parts the following expressions:
\begin{equation}
\label{f26}
\text{Im} A^{WW}_{Z\gamma}=\frac{\alpha}{12}\cot{\theta_W}\sqrt{\frac{t-4M_W^2}{t}}\left(\frac{19}{2}t+16M_W^2\right),
\end{equation}
\begin{equation}
\label{f27}
\text{Im} B^{WW}_{Z\gamma}=-\frac{\alpha}{8}\cot{\theta_W}\sqrt{\frac{t-4M_W^2}{t}}\left(t-4M_W^2\right),
\end{equation}
where the index $WW$ denotes the contribution only from the $W$-boson loop.

Calculations for the other loop amplitudes in Fig.~\ref{fig1ae} with loops of $W$ bosons and gauge bosons, 
as well as ghost fields, are performed using relations \eqref{f10}-\eqref{f11}. Their total contribution 
to $\text{Im} A_{Z\gamma}$, $\text{Im} B_{Z\gamma}$ has the form (the contribution with fermion loops 
is considered separately and given above in \eqref{f17}):
\begin{equation}
\label{f28}
\text{Im} A_{Z\gamma}=\frac{\alpha}{2}\sqrt{\frac{t-4M_W^2}{t}}\left[\frac{t}{12}(19\cot{\theta_W}+
\tan{\theta_W})+\frac{2}{3}M_W^2(4\cot{\theta_W}+\tan{\theta_W})\right],
\end{equation}
\begin{equation}
\label{f29}
\text{Im} B_{Z\gamma}=\frac{\alpha}{2}\sqrt{\frac{t-4M_W^2}{t}}\left[-\frac{t}{2}\cot{\theta_W}+
M_W^2(\cot{\theta_W}-\tan{\theta_W})\right].
\end{equation}
Here, it is useful to emphasize that, in contrast to expressions \eqref{f11e}, the polarization operator of the interaction $Z\gamma$ is not transverse, i.e., the function $B_{Z\gamma}(t)\neq 0$. To construct the hyperfine 
structure potential, which corresponds to vector interaction, we use a dispersion relation for the function 
$A_{Z\gamma}(t)$ of the form \eqref{f18}. This potential will have the same form as \eqref{f14}, but the form 
of the spectral function, denoted below as $\tilde\rho_{Z\gamma}$, will change:
\begin{equation}
\label{f30}
\Delta U^{hfs,V}_{Z\gamma,~W, \omega, c-\text{loops}}(r)=-\frac{2\alpha^2}{9\pi m_1m_2}\frac{\rho_1(1-4\sin^2\theta_W)}{\sin 2\theta_W}
({\bf s}_1{\bf s}_2)
\int\limits_1^\infty\tilde\rho_{Z\gamma}(\xi)\Bigl[\pi\delta({\bf r})-\frac{M_W^2\xi^2}{r}e^{-2M_W\xi r}
\Bigr]d\xi,
\end{equation}
\begin{equation}
\label{f31}
\tilde\rho_{Z\gamma}(\xi)=\frac{\sqrt{\xi^2-1}}{\xi^2}\frac{(\xi^2+\frac{2\rho_2)}{\rho_1}}{(\xi^2-\kappa^2)},
\rho_1=(19 \cot{\theta_W}+\tan{\theta_W}), \rho_2=(4 \cot{\theta_W}+\tan{\theta_W}), \kappa^2=\frac{M_Z^2}{4M_W^2}.
\end{equation}

Calculating the integrals over the coordinates and the spectral parameter $\xi$ using \eqref{f31}, we obtain the following result:
\begin{equation}
\label{f32}
\Delta E^{hfs,V}_{Z\gamma,~W, \omega, c-\text{loops}}=-E_F\frac{\alpha^2\mu \rho_1}{24M_W\kappa^4}
\frac{(1-4\sin^2\theta_W)}{\sin 2\theta_W}\times
\end{equation}
\begin{displaymath}
\Bigl[\kappa^2\Bigl(2\sqrt{1-\kappa^2}+\frac{2\rho_2}{\rho_1}-2\Bigr)+4\frac{\rho_2}{\rho_1}
(\sqrt{1-\kappa^2}-1)\Bigr]=0.3586~\text{Hz}.
\end{displaymath}

The numerical value of this contribution is determined by the appearance of two small parameters 
of the fine structure constant $\alpha$ and the ratio $\frac{\mu}{ M_W}$.

As follows from \eqref{f5a}-\eqref{f5b}, the main contribution of the weak interaction is determined 
by the amplitude of the
single-boson exchange. Therefore, it is also necessary to consider one-loop corrections to this amplitude. 
One of the vertex amplitudes in Fig.~\ref{fig1}(d) is the amplitude in which the $Z$ boson is transformed 
into a pair of $W$ bosons. We calculate the vertex function in such a one-loop amplitude using 
the dispersion approach. The general expression for the interaction amplitude shown in Fig.~\ref{fig1}(d) is:
\begin{equation}
\label{f33}
{\cal M}_{ZWW}=\frac{\alpha^2 \cot\theta_W}{8\pi^2\sin^2\theta_W}\int d^4r 
[\bar u(q_1)\gamma^\nu(1-\gamma_5)\frac{(\hat p_1-\hat r+m_\nu)}{(p_1-r)^2-m_\nu^2}\gamma^\mu(1-\gamma_5) u(p_1)]
[\bar v(p_2)\Gamma^\lambda v(q_2)]\times
\end{equation}
\begin{displaymath}
[g_{\alpha\beta}(2r+k)_\sigma-g_{\beta\sigma}(2k+r)_\alpha+g_{\alpha\sigma}(k-r)_\beta]D_W^{\beta\mu}(r)
D_W^{\alpha\nu}(r+k)\frac{g^{\lambda\sigma}}{(k^2-M_Z^2)}.
\end{displaymath}

To calculate the loop vertex function $\Gamma_{ZWW}^\sigma$ in the leading approximation, we leave only the 
loop momentum $r$ in the numerator and neglect the exchange momentum $k$. Then in the numerator of this function 
we obtain:
\begin{equation}
\label{f34}
{\cal N}^\sigma_{ZWW}=[g_{\alpha\beta}(2r+k)_\sigma-g_{\beta\sigma}(2k+r)_\alpha+g_{\alpha\sigma}(k-r)_\beta]
\gamma^\alpha(1-\gamma_5)(\hat p_1-\hat r)\gamma^\beta(1-\gamma_5)=
\end{equation}
\begin{displaymath}
8r^\sigma\hat r(1-\gamma_5)+4r^2\gamma^\sigma(1-\gamma_5)=6r^2\gamma^\sigma(1-\gamma_5).
\end{displaymath}

In the dispersion approach, the loop vertex function $\Gamma_{ZWW}^\sigma$ is defined by the dispersion integral 
with one subtraction:
\begin{equation}
\label{f35}
\Gamma^\sigma_{ZWW}=\gamma^\sigma(1-\gamma_5)F_{ZWW}(t),~
F_{ZWW}(t)=\frac{(t-M_Z^2)}{\pi}\int_{4M_W^2}^\infty\frac{Im F_{ZWW}(t')dt'}{(t'-M_Z^2)(t'-t-i0)},
\end{equation}
\begin{equation}
\label{f36}
Im F_{ZWW}(t)=\frac{3}{8\pi}\sqrt{\frac{t-4M_W^2}{t}}.
\end{equation}
At $t=M_Z^2$ the interaction between the $Z$ boson and the lepton is determined by the vertex function \eqref{f1}.

Given the vertex function \eqref{f36}, we can construct the hyperfine potential of the electron-muon interaction 
from the amplitude \eqref{f33}. As in the case of the single-boson interaction \eqref{f5}, it can be divided into 
two parts, determined by the vector and pseudovector currents.
In the momentum representation, the contribution of the vector current to the hyperfine part of the interaction 
operator is determined using \eqref{f3}-\eqref{f3c} by the following expression:
\begin{equation}
\label{f37}
\Delta U^{hfs}_{V,ZWW}({\bf k})=\frac{\alpha^2 \cot^2\theta_W(1-4\sin^2\theta_W)}{m_1m_2 \sin 2\theta_W}
\frac{{\boldsymbol\sigma}_1{\boldsymbol\sigma}_2}{4}\int_1^\infty \frac{\sqrt{\xi^2-1}d\xi}{(\xi^2-\frac{M_Z^2}{4M_W^2})}
\left[
1-\frac{4\xi^2M_W^2}{{\bf k}^2+4M_W^2\xi^2}
\right].
\end{equation}

After the Fourier transform of the potential \eqref{f37}, we obtain a potential in coordinate representation, 
which is a superposition of the $\delta$-potential and the Yukawa-type potential. Averaging it 
on the ground-state wave functions and integrating on the spectral parameter $\xi$, we find the final 
expression for the contribution \eqref{f37} to the hyperfine splitting in the form:
\begin{equation}
\label{f38}
\Delta E^{hfs}_{V,ZWW}=E_F
\frac{3\mu M_W\alpha^2 \cot^2\theta_W(1-4\sin^2\theta_W)}{2\sin 2\theta_W M^2_Z}
\left(1-\sqrt{1-\frac{M_Z^2}{4M_W^2}}\right)=0.1392~Hz.
\end{equation}

The pseudovector current in \eqref{f35} also contributes to the hyperfine structure. Using the corresponding 
bispinor convolutions \eqref{f3}-\eqref{f3c}, we obtain its contribution to the interaction operator 
in the momentum representation:
\begin{equation}
\label{f39}
\Delta U^{hfs}_{AV,ZWW}({\bf k})=\frac{6\alpha^2 \cot^2\theta_W}{\sin 2\theta_W}
\frac{{\boldsymbol\sigma}_1{\boldsymbol\sigma}_2}{4}\int_1^\infty \frac{\sqrt{\xi^2-1}d\xi}{(\xi^2-\frac{M_Z^2}{4M_W^2})}
\frac{1}{({\bf k}^2+4M_W^2\xi^2)}.
\end{equation}

In the coordinate representation, the corresponding interaction operator is equal to
\begin{equation}
\label{f40}
\Delta U^{hfs}_{AV,ZWW}({\bf r})=\frac{3\alpha^2 \cot^2\theta_W}{2\pi\sin 2\theta_W}
\frac{{\boldsymbol\sigma}_1{\boldsymbol\sigma}_2}{4}\int_1^\infty \frac{\sqrt{\xi^2-1}d\xi}
{(\xi^2-\frac{M_Z^2}{4M_W^2})}\frac{1}{r}e^{-2M_W\xi r}.
\end{equation}

By calculating the matrix elements of the potential \eqref{f40}, we obtain the contribution 
to the hyperfine structure of muonium:
\begin{equation}
\label{f41}
\Delta E^{hfs}_{AV,ZWW}=E_F
\frac{9\alpha m_1m_2 M_W \cot^2\theta_W}{2\pi\sin 2\theta_W M^3_Z}
\left[\frac{M_Z}{2M_W}-\sqrt{1-\frac{M_Z^2}{4M_W^2}}\arcsin\bigl(\frac{M_Z}{2M_W}\bigr)\right]=0.0786~\text{Hz}.
\end{equation}

Along with the contribution of the $WW$-boson loop in Fig.~\ref{fig1}(d) to the vertex function, there are also contributions from the unphysical gauge bosons $\omega^{\pm}$. However, their contribution, which is suppressed 
by the vertex mass factor $m_e/M_W$, $m_\mu/M_W$ from the vertex of interaction of $\omega^{\pm}$ with leptons, 
can be neglected compared to \eqref{f34}.

In conclusion of this section, we note that the interaction amplitude shown in Fig.~\ref{fig1}(b) is accounted 
for in the muonium hyperfine structure when using the experimental value of the electron (muon) anomalous 
magnetic moment \cite{brodsky}.

\section{Contribution of Box-Type $Z\gamma$, $ZZ$ Interaction Amplitude}

\begin{figure}[htbp]
\centering
\includegraphics[scale=1.]{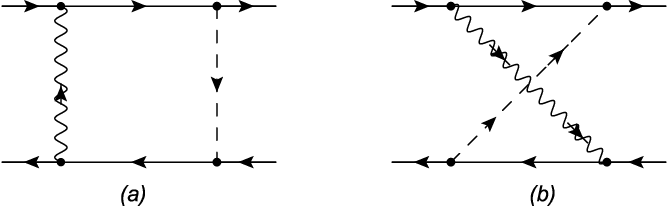}
\caption{Amplitude of photon and $Z$-boson interaction. The wavy line represents the photon and the dotted line represents the $Z$ boson.}
\label{fig2}
\end{figure}

Box-type amplitudes with different exchange bosons have already been studied for various physical processes in \cite{alcorta,weak3,weak4,weak5a,weak5,weak6}. 
The box-type amplitudes we consider in this section are determined by finite integrals, so to calculate them, 
we use not the dispersion method but a direct calculation of the four-dimensional integral.
We use here the same calculation technique as in the case of two-photon exchange amplitudes \cite{apm2018}.
Let us first consider the direct and crossed interaction amplitudes of the muon and electron 
in the case of photon and $Z$-boson exchange, shown in Fig.~\ref{fig2}. Taking into account 
the vertex factors that determine the interaction of photons and $Z$ bosons with leptons, 
we can represent the direct box-type amplitude as:
\begin{equation}
\label{e1}
{\cal M}^{\Box \gamma Z}_{\text{dir}}=\frac{(4\pi\alpha)^2}{\sin^22\theta_W}
\int\frac{d^4 k}{(2\pi)^4}\left[\bar u(q_1)\Gamma^\nu\frac{(\hat p_1-\hat k+m_1)}
{(p_1-k)^2-m_1^2}\gamma^\mu u(p_1)\right]D_Z^{\mu\lambda}(k)D^{\nu\sigma}(k)\times
\end{equation}
\begin{displaymath}
\left[\bar v(p_2)\Gamma^\lambda\frac{(-\hat p_2-\hat k+m_2)}
{(p_2+k)^2-m_2^2}\gamma^\sigma v(q_2)\right],
\end{displaymath}
where, as before, the $Z$-boson propagator is chosen in the Coulomb gauge.
From here on, the amplitudes and contributions of the box type are indicated by the symbol $\Box$.
Then, using the transformation law of Dirac bispinors when going to the rest frame, and introducing 
projection operators into states of the muon-electron pair with spin 1 and 0, we obtain 
the following expression for the amplitude \eqref{e1}:
\begin{equation}
\label{e2}
{\cal M}^{\Box \gamma Z}_{\text{dir}}=\frac{\alpha^2}{\pi^2\sin^22\theta_W}
\int d^4k \text{Tr}\left[\hat\Pi^\ast\Gamma^\nu\frac{(\hat p_1-\hat k+m_1)}{(k^2-2m_1k_0)}\gamma^\mu
\hat\Pi\Gamma^\lambda\frac{(-\hat p_2-\hat k+m_2)}{(k^2+2m_2 k_0)}
\gamma^\sigma\right]\times
\end{equation}
\begin{displaymath}
\frac{1}{(k^2-m_Z^2)}\left[g^{\mu\lambda}+\frac{1}{2({\bf k}^2+m_Z^2)}(f^\mu k^\lambda+f^\lambda k^\mu)\right]
D^{\nu\sigma}(k),
\end{displaymath}
\begin{equation}
\label{e2a}
\hat\Pi_{S=0,1}=\frac{(1+\gamma_0)}{2\sqrt{2}}\hat\varepsilon(\gamma_5),
\end{equation}
where $\hat\varepsilon$ is the spin polarization vector.
For the state with spin 0, one must use the replacement $\hat\varepsilon\to \gamma_5$.

The crossed interaction amplitude has a similar structure and can be represented as follows:
\begin{equation}
\label{e3}
{\cal M}^{\Box \gamma Z}_{\text{cr}}=\frac{\alpha^2}{\pi^2\sin^22\theta_W}
\int d^4k \text{Tr}\left[
\hat\Pi^\ast\Gamma^\nu\frac{(\hat p_1-\hat k+m_1)}{(k^2-2m_1k_0)}\gamma^\mu
\hat\Pi\gamma^\sigma\frac{(-\hat p_2+\hat k+m_2)}{(k^2-2m_2 k_0)}
\Gamma^\lambda\right]\times
\end{equation}
\begin{displaymath}
\frac{1}{(k^2-m_Z^2)}\left[g^{\mu\lambda}+\frac{1}{2({\bf k}^2+m_Z^2)}(f^\mu k^\lambda+f^\lambda k^\mu)\right]
D^{\nu\sigma}(k).
\end{displaymath}

Further calculations of the trace and different substitutions are conveniently performed using 
the Form package \cite{Form}. As a result, the total contribution of the two amplitudes 
to the particle interaction operator can be represented as an integral over the loop momentum:
\begin{equation}
\label{e4}
\Delta V^{hfs}_{\Box Z\gamma}=\frac{\alpha^2}{3\pi^2\sin^22\theta_W}
\int \frac{d^4k}{k^2(k^2-m_Z^2)(k^4-4m_1^2k_0^2)(k^4-4m_2^2k_0^2)}\times
\end{equation}
\begin{displaymath}
\Bigl[2m_1m_2k_0^4+2m_1m_2k_0^4\frac{k^2}{({\bf k}^2+m_Z^2)}-\frac{1}{2}k^4k_0^2(1-2a_z)^2-m_1m_2
\frac{k^4k_0^2}{({\bf k}^2+m_Z^2)}-k^6(1-2a_z)^2\Bigr].
\end{displaymath}

To calculate the momentum integrals, we move to Euclidean space by replacing $k_0\to ik_0$, $k^2\to -k^2$:
\begin{equation}
\label{e5}
\int d^4 k=4\pi\int k^3 dk \int_0^\pi \sin^2\phi d\phi,\quad k_0=k\cos\phi,\quad |{\bf k}|=k\sin\phi.
\end{equation}

Then the integral \eqref{e4} splits into a sum of integrals, each of which can be evaluated analytically:
\begin{equation}
\label{e6}
I_1=\int_0^\infty k dk\int_{-1}^1 \frac{x^2\sqrt{1-x^2} dx}{(k(1-x^2)+1)(k+1)(k+a_1^2x^2)(k+a_2^2x^2)}=
\end{equation}
\begin{displaymath}
\frac{\pi}{64(a_1^2-a_2^2)}\left[11a_1^2a_2^2\ln\frac{a_1}{a_2}+8a_2^2(2\ln 4a_2-3)-8a_1^2(2\ln 4a_1-3)\right],
\end{displaymath}
where we have introduced the notation: $a_1=2m_1/m_Z$, $a_2=2m_2/m_Z$. The result in square brackets 
\eqref{e6} is presented as an expansion in $a_1$, $a_2$ up to second order. Other integrals are of the form:
\begin{equation}
\label{e7}
I_2=\int_0^\infty k dk\int_{-1}^1 \frac{x^6\sqrt{1-x^2} dx}{(k(1-x^2)+1)(k+1)(k+a_1^2x^2)(k+a_2^2x^2)}=
\end{equation}
\begin{displaymath}
\frac{\pi}{1536(a_1^2-a_2^2)}\left[15a_1^2(7a_2^2-8)\ln a_1-15a_2^2(7a_1^2-8)\ln a_2+2(24\ln 2-19)(a_1^2-a_2^2)\right],
\end{displaymath}
\begin{equation}
\label{e8}
I_3=\int_0^\infty k dk\int_{-1}^1 \frac{x^2\sqrt{1-x^2} dx}{(k+1)(k+a_1^2x^2)(k+a_2^2x^2)}=
\end{equation}
\begin{displaymath}
\frac{\pi}{16(a_1^2-a_2^2)}\left[(a_1^2-a_2^2)(4\ln 2-1)+4a_2^2\ln a_2-4a_1^2\ln a_1\right],
\end{displaymath}
\begin{equation}
\label{e9}
I_4=\int_0^\infty k dk\int_{-1}^1 \frac{\sqrt{1-x^2} dx}{(k+1)(k+a_1^2x^2)(k+a_2^2x^2)}=
\end{equation}
\begin{displaymath}
\frac{\pi}{2(a_1^2-a_2^2)}\left[a_1^2(2\ln 2+1-2\ln a_1)-a_2^2(2\ln 2+1-2\ln a_2)\right],
\end{displaymath}
\begin{equation}
\label{e10}
I_5=\int_0^\infty k dk\int_{-1}^1 \frac{x^4\sqrt{1-x^2} dx}{(k(1-x^2)+1)(k+1)(k+a_1^2x^2)(k+a_2^2x^2)}=
\end{equation}
\begin{displaymath}
-\frac{\pi}{256(a_1^2-a_2^2)}\left[(a_1^2-a_2^2)(24\ln 2-7)+64\ln\frac{a_1}{a_2}-44(a_1^2\ln a_2-a_2^2\ln a_1)\right].
\end{displaymath}

Then the total contribution to the hyperfine splitting of the $1S$ state in muonium is determined 
by the sum of the integral terms \eqref{e6}-\eqref{e10}:
\begin{equation}
\label{e11}
\delta E^{hfs}_{\Box Z\gamma}=E_F\frac{m_1m_2\alpha}{8\pi m_Z^2\sin^22\theta_W (a_2^2-a_1^2)}
\Biggl\{4a_1a_2\ln\frac{a_1}{a_2}+
\end{equation}
\begin{displaymath}
(1-2a_z)^2\left[(15+36\ln 2)(a_1^2-a_2^2)+
36a_2^2\ln a_2-36a_1^2\ln a_1\right]\Biggr\}.
\end{displaymath}

Compared to the leading contribution, this contribution is suppressed by the additional $\alpha$ 
power and the second-order mass ratio.
The numerical value of the contribution for the ground state of muonium is equal to
\begin{equation}
\label{e12}
\Delta E^{hfs}_{\Box Z\gamma}=-0.0380~\text{Hz}.
\end{equation}

By replacing the photon with another $Z$ boson in the interaction amplitudes in Fig.~\ref{fig2}, 
one can obtain a contribution to the corresponding box interaction operator in the form:
\begin{equation}
\label{e12a}
\Delta V^{hfs}_{\Box ZZ}=\frac{4\alpha^2(1-2a_z+2a_z^2)^2}
{3\pi^2\sin^42\theta_W}\int\frac{k^4(k_0^2+2k^2) d^4k}{(k^2-M_Z^2)^2(k^4-4m_1^2k_0^2)(k^4-4m_2^2k_0^2)},
\end{equation}
where in the numerator we left the leading term for $m_1$, $m_2$. Further integration in 4-dimensional 
Euclidean space is carried out similarly to \eqref{e6}-\eqref{e10}. The result of integration in next 
to leading order with respect to $a_1$, $a_2$ leads to a contribution to hyperfine splitting:
\begin{equation}
\label{e12b}
\Delta E^{hfs}_{\Box ZZ}=-E_F\frac{9\alpha m_1m_2(1-2a_z+2a_z^2)^2}{8\pi\sin^42\theta_W M_Z^2}
\Bigl[1+\frac{15(a_1^4\ln\frac{a_1}{2}-a_2^4\ln\frac{a_2}{2})}
{32(a_1^2-a_2^2)}+\frac{47(a_1^2+a_2^2)}{108}\Bigr]
=-0.0403~\text{Hz}.
\end{equation}

The obtained results \eqref{e12} and \eqref{e12b} are of the same order.

\begin{figure}[htbp]
\centering
\includegraphics[scale=1.2]{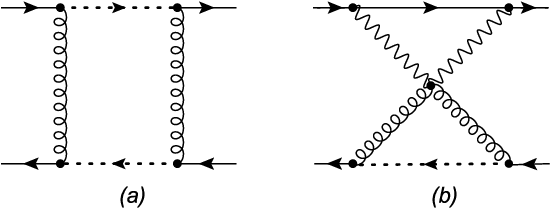}
\caption{$W$-boson interaction amplitudes. The dashed line in the intermediate state denotes a neutrino.}
\label{fig3}
\end{figure}

\section{Contribution of Box-Type WW Interaction Amplitude}

The contribution to the muon-electron interaction operator is determined by another "box" amplitude 
with two $W$ bosons, shown in Fig.~\ref{fig3}(a). The original expression for this amplitude is the following:
\begin{equation}
\label{e13}
{\cal M}^\Box_{WW}=\frac{\pi^2\alpha^2}{4\sin^42\theta_W}
\int\frac{d^4 k}{(2\pi)^4}\left[\bar u(q_1)\gamma^\nu(1-\gamma_5)\frac{(\hat p_1-\hat k+m_{\nu_e})}
{(p_1-k)^2-m_{\nu_e}^2}\gamma^\mu(1-\gamma_5) u(p_1)\right]\times
\end{equation}
\begin{displaymath}
\left[\bar v(p_2)\gamma^\lambda(1-\gamma_5)\frac{(-\hat p_2-\hat k+m_{\nu_\mu})}
{(p_2+k)^2-m_{\nu_\mu}^2}\gamma^\sigma (1-\gamma_5)v(q_2)\right]D_W^{\mu\lambda}(k)D_W^{\nu\sigma}(k).
\end{displaymath}

The peculiarity of this interaction amplitude is that there is a pair of neutrinos in the intermediate 
state. Transformations of this amplitude are performed in a way similar to the transformations of \eqref{e1}. 
Again, introducing projection operators onto states of the electron-muon pair with definite spin 
and calculating the total trace of the Dirac factors, we represent it firstly in the integral form:
\begin{equation}
\label{e14}
{\cal M}^\Box_{WW}=\frac{\alpha^2}{48\pi^2 M_W^2\sin^42\theta_W}
\int \frac{d^4k}{((k^2-\frac{a_1^2}{4})^2+a_1^2k_0^2)((k^2-\frac{a_2^2}{4})^2+a_2^2k_0^2)(k^2+1)^2
({\bf k}^2+1)^2}\times
\end{equation}
\begin{displaymath}
\Bigl\{
16a_1a_2k_0^4-32a_1a_2k_0^6+16a_1a_2k_0^8-4k^2k_0^2(a_1^2+a_2^2+4a_1a_2)+32a_1a_2k^2k_0^4-16a_1a_2k^2k_0^6+
\end{displaymath}
\begin{displaymath}
k^4k_0^2(16-8a_1^2-8a_2^2-4a_1a_2)-k^4k_0^4(32-2a_1^2-2a_2^2+10a_1a_2)+16k^4k_0^6+4k^4(a_1^2+a_2^2)+
\end{displaymath}
\begin{displaymath}
k^6k_0^2(16-a_1^2-a_2^2+12a_1a_2)-8k^6k_0^4-k^6(16-2a_1^2-2a_2^2+2a_1a_2)-12k^8k_0^2-k^8(8+(a_1+a_2)^2)+
4k^{10}\Bigr\},
\end{displaymath}
where we have also moved to Euclidean space when integrating over $k$. When computing this amplitude in the leading 
order in $a_1$, $a_2$, one can immediately set $a_1=a_2=0$ in both the numerator and the denominator. 
Then the expression \eqref{e14} is significantly simplified and takes the form:
\begin{equation}
\label{e15}
{\cal M}^\Box_{WW}=\frac{\alpha^2}{24\pi^2 M_W^2\sin^42\theta_W}
\int \frac{d^4k}{k^4(k^2+1)^2({\bf k}^2+1)^2}
\Bigl\{4k_0^2-8k_0^4+4k_0^6+4k^2k_0^2-2k^2k_0^4-
\end{equation}
\begin{displaymath}
4k^2-3k^4k_0^2-2k^4+k^{6}\Bigr\}.
\end{displaymath}

Analytical integration in \eqref{e15} over the loop momentum yields the following result:
\begin{equation}
\label{e16}
\Delta E^{hfs}_{\Box WW}=-\frac{\mu^3\alpha^5\left(32\ln 2-\frac{361}{15}\right)}{12\pi M_W^2\sin^42\theta_W}.
\end{equation}

This contribution is fifth order of smallness in the fine-structure constant $\alpha$, but the presence of the square of the $W$-boson mass in the denominator leads to a significant reduction in the numerical value, which for the ground state is equal
\begin{equation}
\label{e17}
\Delta E^{hfs}_{\Box WW}=-0.1026~\text{Hz}.
\end{equation}

We do not consider other amplitudes of the box type with two $Z$ bosons, since, according to our estimates, 
their contribution can be neglected at this stage of research.
The amplitude in Fig.~\ref{fig3}(b) does not contribute to the hyperfine structure spectrum 
in the leading order in $a_1$, $a_2$ and is therefore negligibly small.

\section{Conclusion}

A new, high-precision measurement of the muonium hyperfine structure was performed at J-PARC MLF MUSE using a pulsed muon beam. This measurement is an intermediate step toward future, more precise measurements with further increases in beam intensity. The following result was obtained: $\nu_{HFS}$ = 4463.302(4) MHz (0.9 ppm) \cite{kanda}.
This result was consistent with the previous ones obtained at Los Alamos National Laboratory \cite{paper10a} and the current theoretical calculation \cite{paper1}.
To determine the resonant frequency, Rabi oscillation spectroscopy was developed in \cite{nishimura}. The analysis yielded a hyperfine splitting value of $\nu_{HFS}$ = 4463301.61 (0.71) kHz, which is somewhat more accurate and also consistent with previous experimental data.
The expected significant progress in the experimental study of hyperfine splitting in muonium will bring this problem to a new level of research into subtle effects in the physics of simple atomic systems, caused not only by corrections to the electromagnetic interaction but also by the effects of the electroweak and strong interactions \cite{eides1,eides2,egs1,sgk}.

This work investigates electroweak corrections to the muonium hyperfine structure. Although these corrections are small and have been neglected until now in the determination of the complete theoretical expression for the hyperfine splitting of the ground state in muonium, the growing experimental accuracy at the current stage \cite{paper2,paper10} requires accounting for contributions on the order of 1 Hz. The study of such corrections is also of purely theoretical interest.
The calculation of electroweak radiative corrections is important in many physical problems, and their study has been the subject of many works \cite{alcorta,weak3,weak4,weak5a,weak5,weak6}. In the case of the muonium hyperfine structure, radiative corrections were studied in \cite{weak7} using dimensional regularization and renormalization techniques.

In our work, the dispersion method of renormalization is used to calculate one-loop corrections to particle propagators, and each considered interaction amplitude is represented by a separate expression when constructing the particle interaction operator and its contribution to the energy spectrum.
When calculating the amplitudes of box-type interactions, which are determined by finite integral expressions, we use the usual approach of going to the Euclidean space and directly integrating over the corresponding variables.

The total electroweak contribution to the hyperfine splitting in muonium is determined by the sum of the calculated corrections \eqref{f5a}, \eqref{f5b}, \eqref{f15}, \eqref{f23a}, \eqref{f32}, \eqref{f38}, \eqref{f41}, \eqref{e12}, \eqref{e12b}, \eqref{e17} and is equal to
\begin{equation}
\label{e18}
\Delta E^{hfs}_{tot}(1S)=-70.1217~\text{Hz}.
\end{equation}

It is worth noting here once again that when calculating electroweak corrections, we present the results to four decimal places and maintain the same notation for the full result \eqref{e18}. Some unaccounted corrections may contribute on the order of $10^{-4}\div 10^{-3}$ Hz.
Most of the result \eqref{e18} is determined by the amplitude of the one $Z$-boson exchange, but a number of propagator-type corrections add up to several Hz.

A detailed comparison of individual contributions to the hyperfine structure of muonium with previous calculations is quite difficult to perform, since the existing work \cite{weak7} on this problem presents the total contribution in a summary numerical form.
Nevertheless, it is possible to outline the numerical results obtained in the work \cite{weak7}:
$(-160 + (-2.8 + 0.4) 10^{-2})$ mHZ,
where three numbers refer to major contributions from box, self-energy and vertex diagrams.
It should be noted that our obtained box amplitude contribution (-0.1809~Hz) agrees with these results, taking into account that we consider only three box diagrams.
The main contribution in our calculations is due to corrections related to vacuum polarization effects in $\gamma\gamma$- and $Z\gamma$ interactions. Their total value amounts to several hertz and differs significantly from the results of the study \cite{weak7}. Our total contribution to hyperfine splitting in muonium may be useful for comparison with new experimental data.

\begin{acknowledgments}

This work is supported by the Russian Science Foundation (Grant No. 25-72-00029) (F.A.M.).
\end{acknowledgments}



\end{document}